\documentclass[letterpaper]{article}
\pdfoutput=1
\usepackage{graphicx}
\usepackage{amsmath}
\begin{document}

\begin{center}
\LARGE {Probing the statistic in the cosmic microwave background}
\linebreak

\small{T.Ghahramanyan, S.Mirzoyan, E.Poghosian, G.Yegorian}

\end{center}

\small{Yerevan Physics Institute and Yerevan State University, Yerevan, Armenia}

\begin{abstract}
Kolmogorov's statistic is used for the analysis of properties of perturbations in the Cosmic Microwave Background signal. 
We obtain the maps of the Kolmogorov stochasticity parameter for W and V band temperature data of WMAP which are differently affected by the Galactic disk radiation and then we model datasets with various statistic of perturbations.  The analysis shows that the Kolmogorov's parameter can be an efficient tool for the separation of Cosmic Microwave Background from the contaminating radiations due to their different statistical properties.
\end{abstract}

\section{Introduction}

The maps of Cosmic Microwave Background (CMB) radiation are among important sources of cosmological information.
Various descriptors have been applied to study the non-Gaussianities in the CMB temperature maps. Kolmogorov stochasticity parameter (K-parameter) 
[\ref{Kolm},\ref{Arnold},\ref{Arnold1}] when applied to CMB temperature datasets[\ref{GK_KSP},\ref{K_sky}] reveals the Galactic disk, hence, outlining the differences in the statistical properties of the disk radiation and that of CMB. The resulting Kolmogorov's map shows enhanced randomness, i.e. higher value of K-parameter in certain regions including the Cold Spot [\ref{Cruz}], the non-Gaussian anomaly in the southern sky. Notably, the K-parameter of the Cold Spot satisfies the criterion expected for the voids, namely, increase of the randomness towards the walls of the void which act as hyperbolic (deviating) lenses for the CMB [\ref{GK_void}, \ref{Das_Sp}].

The present paper is devoted to the analysis of the interplay between the properties of datasets, which are crucial at separation of signals of different stochasticity (randomness) characteristics. First, we obtain the K-map for W and V bands of Wilkinson Microwave Anisotropy Probe (WMAP) data, which indicate the differences in the statistic at different contribution of the Galactic radiation. We then estimate the K-parameter for datasets of Gaussian, along with a perturbed Gaussian distributions. We start with the definition of the Kolmogorov statistic [\ref{Kolm}].     

\section{Method}

For a finite random sequence of numbers the Kolmogorov's theorem [\ref{Kolm}] aims to determine its stochastic (randomness) properties.
Consider a sequence of real numbers $x_1, x_2, \ldots , x_n$ and that $x_1 \leq x_2 \leq \ldots \leq x_n$. Having a continuous distribution function

\begin{equation}
F(X) = P\{x \leq X\}
\label{eq:distribution}
\end{equation}
where $P$ is the probability of event in brackets. The empirical distribution function is defined as

\begin{eqnarray*}
F_n(X)=
\begin{cases}
0\ , & X<x_1\ ;\\
k/n\ , & x_k\le X<x_{k+1},\ \ k=1,2,\dots,n-1\ ;\\
1\ , & x_n\le X\ .
\end{cases}
\label{eq:empiricdistribution}
\end{eqnarray*}

Kolmogorov's stochasticity parameter is defined as

\begin{equation}
\lambda_n = \sqrt{n}\, sup|F(X)-F_n(X)|
\label{eq:lambda}
\end{equation}
which according to the theorem is a random number defined by a distribution function $\Phi_n(\lambda) = P\{\lambda_n \leq \lambda\}$ that tends to a function $\Phi(\lambda)$ uniformly when $n \rightarrow \infty$, for any continous distribution $F(x)$

\begin{equation}
\Phi_n(\lambda) \rightarrow \Phi(\lambda).
\label{eq:philimit}
\end{equation}
where 

\begin{equation}
\Phi(\lambda) = \sum_{k=-\infty}^{+\infty}{(-1)^k e^{-2k^2\lambda^2}},\\
\Phi(0) = 0.
\label{eq:phi}
\end{equation}

%For small values of $\lambda$ we can use approximation $\Phi(\lambda) \approx \frac{\sqrt(2\pi)}{\lambda} e^{-{\pi}^2/8\lambda^2} $

\begin{figure}[!htbp]
	\centering
		\includegraphics[width=2.3 in]{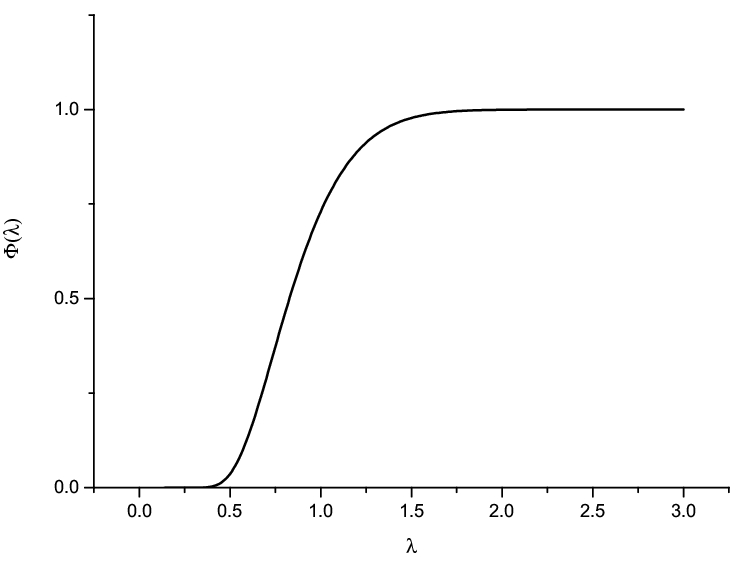}
		\includegraphics[width=2.3 in]{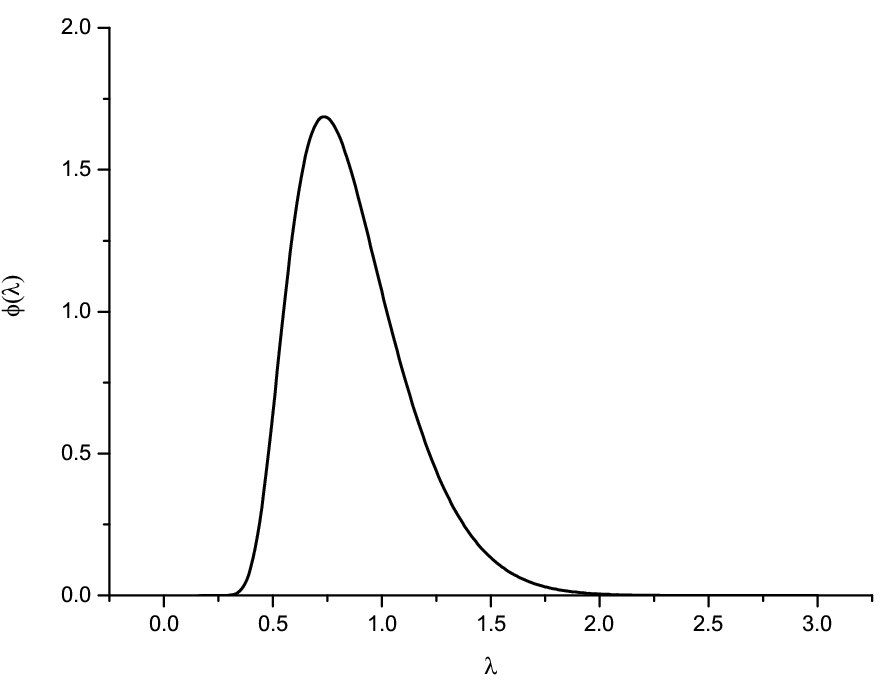}
	\caption{Kolmogorov's distribution $\Phi(\lambda)$ (a) and it's derivative $\phi(\lambda)=\Phi'(\lambda)$ (b).}
	\label{fig:fi}
\end{figure}

From Figure \ref{fig:fi} we see that the most probable values of $\lambda$ are between 0.3 and 2.2.

This means that for big enough $n$ and random sequence $x_n$ the Kolmogorov's stochasticity parameter $\lambda_n$ will have a distribution close to $\Phi(\lambda)$ and if the sequence is not random, the distribution will be different.
This gives us an opportunity to test the randomness of a sequence. 

Note, that for each studied sequence we can only get one value of $\lambda_n$, so that it is not possible to guess whether that number belongs to a certain distribution. Here we have an option to split a sequence into subsets, if the length of the sequence,  $n$, is big enough.

Consider a split of the sequence $x_n$ into $m$ subsequences. We can get a new sequence $\lambda_n^m$ of length $m$, which should be a random sequence with distribution close to $\Phi(\lambda)$.
So $\lambda_n^m$ is our option to proceed on the sequences.

\section{Numerical Analysis}

The K-maps for the W and V bands of the WMAP 5-year data are given in Fig.2. 
\begin{figure}[!htbp]
	\centering
		\includegraphics[width=2.8 in]{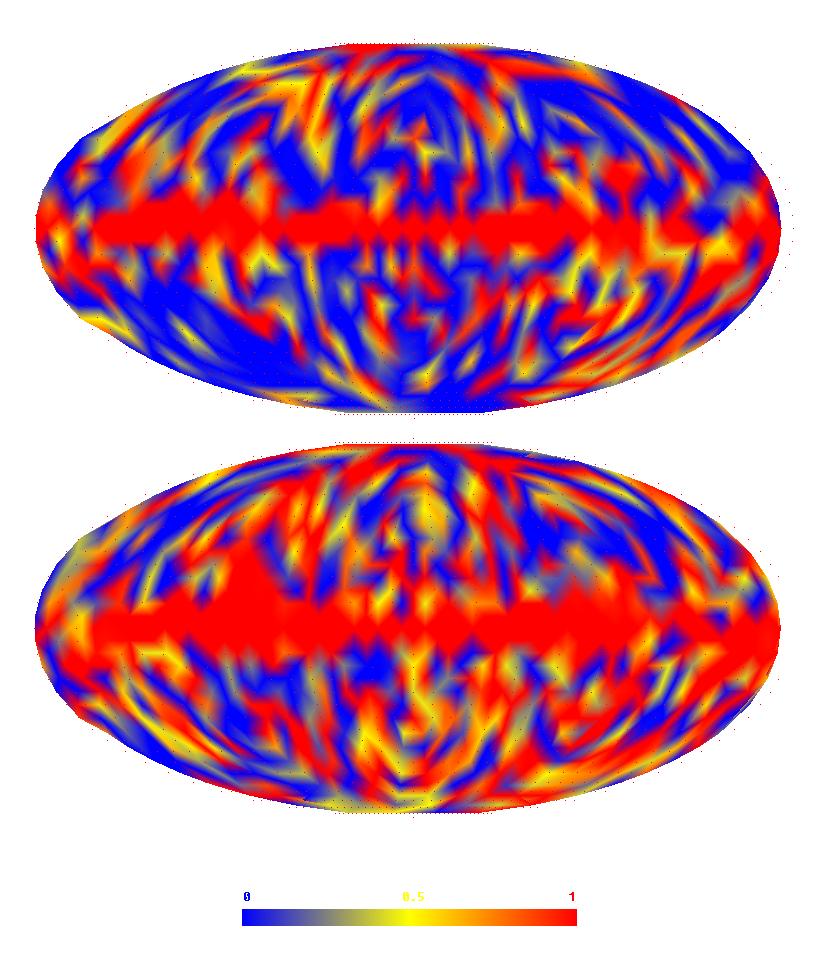}
	\caption{Kolmogorov maps computed for Wilkinson Microwave Anisotropy Probe's W and V band CMB temperature data.}
	\label{fig:fi1}
\end{figure}

The visible differences are due to the different role of the Galactic disk contribution, i.e. of a perturbation.
We will now model such an effect for the K-parameter. Our first aim is to find parameters and ranges for their values that will show whether we are dealing with stochastic sequence or no. The stochastic sequence can be resulted e.g. from a hyperbolic dynamical system [\ref{Arnold_book}] but can also have another origin. The length of the considered sequences are taken as 4096 which corresponds to the CMB maps in given HEALPix format of pixelization.  

We generate sequences of given length with certain properties, namely, we take 400 different sequences that are divided into 4 following groups each having 100 sequence:

1st hundred are random sequences with Gaussian distribution;

2nd hundred are random sequences with perturbed Gaussian distribution;

3rd hundred are sequences with Gaussian distribution, but they are not random but are extracted from a periodic sequence like 
$x_n=a$ $n$ mod $b$, where $a$ and $b$ are constants;

4th hundred are sequences including two copies of the same random Gaussian sequence.

Our choice of Gaussian distribution for $F(x)$  is due to its interest for CMB maps, but in principle, it could be anything including a uniform distribution.

So we have groups 1 and 2 that are stochastic and 3 and 4 which are non stochastic, and we can proceed with calculations for all sequences.
In all figures the horizontal axis $i$ denotes the index of a sequence that runs from $1$ to $400$, i.e. if $1 \leq i < 100$ we have values of the 1st group, if $101 \leq i < 200$ for the 2nd group, and so on.

\begin{figure}[!htbp]
	\centering
		\includegraphics[width=2.3 in]{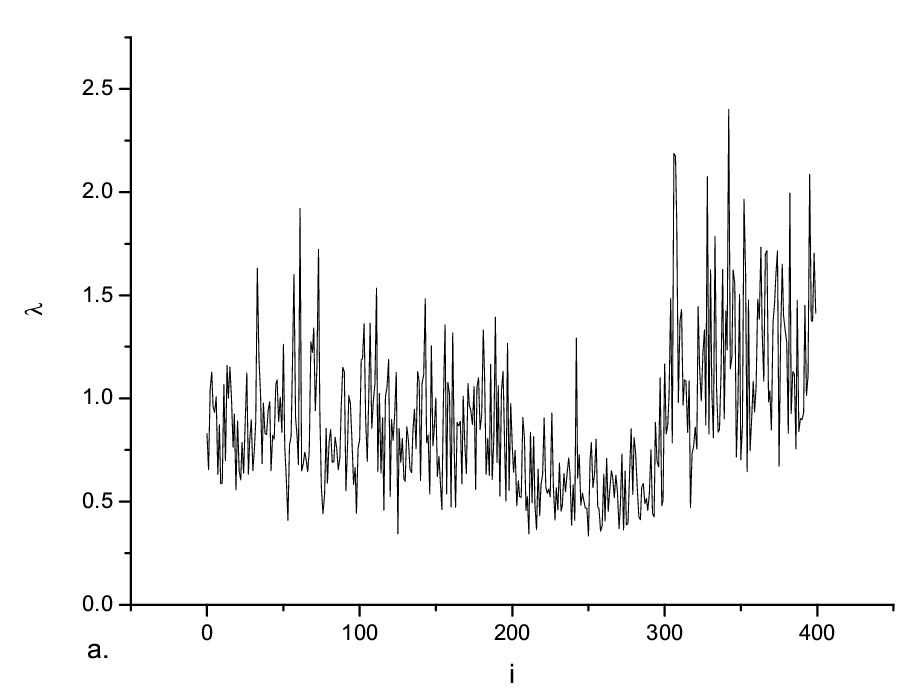}
		\includegraphics[width=2.3 in]{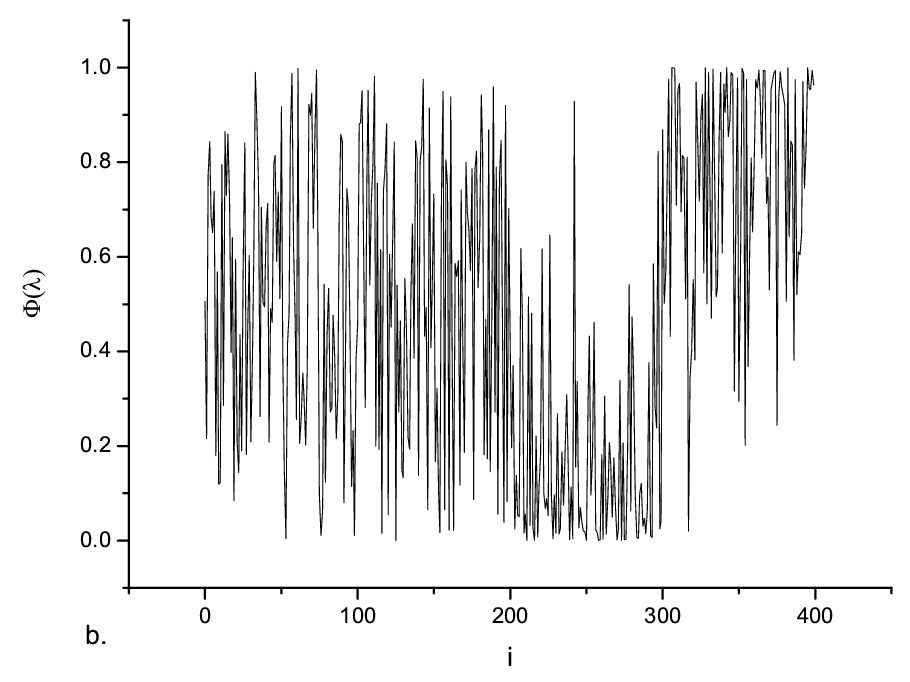}
		\includegraphics[width=2.3 in]{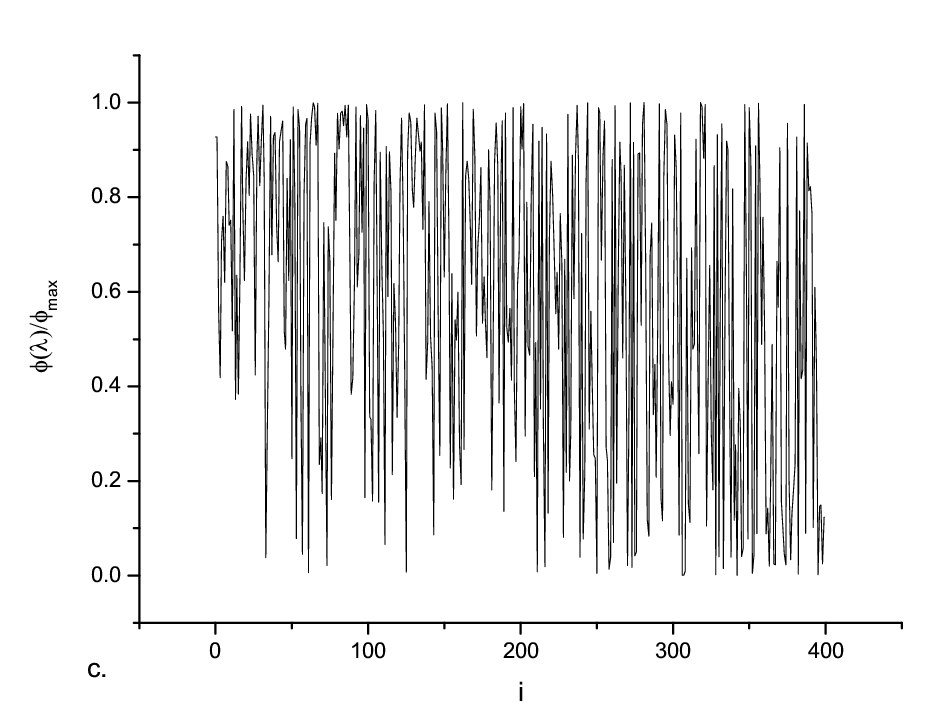}
		\includegraphics[width=2.3 in]{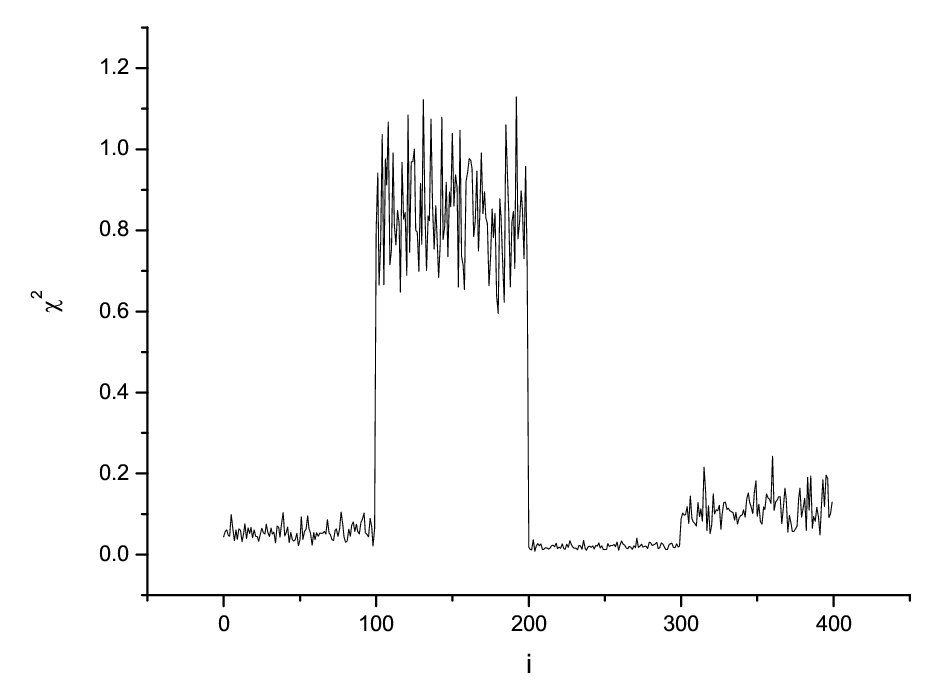}
	\caption{Computed values of $\lambda$ (a), $\Phi(\lambda)$ (b), $\phi(\lambda)/\phi_{max}(\lambda)$ (c) and $\chi^2$ (d).}
	\label{fig:test_1}
\end{figure}

Figure \ref{fig:test_1} shows results of computations.
In the \ref{fig:test_1}.a the values of $\lambda$ are shown for each sequence, we see correspondence to distribution $\Phi(\lambda)$ for 1st and 2nd groups, but not for 3rd or 4th ones.
Figure \ref{fig:test_1}.b contains similar plot for values $\Phi(\lambda)$, in this case 1st and 2nd groups should have almost uniformly distributed values between 0 and 1, but not 3rd or 4th, and this is what we see.
Figure \ref{fig:test_1}.c shows that values $\phi(\lambda)/\phi_{max}(\lambda)$ are close to 1 in 1st and 2nd case, indicating right distribution again.

Finally, \ref{fig:test_1}.d, is the $\chi^2$ calculated for each sequence's distribution as compared with theoretical Gaussian distribution.

One can state that none of these analysis gives precise result/range of values, when it comes to a single sequence. The first 3 sequences do not allow to conclude whether a single sequence at index $i$ is random or not, and the 4th figure of $\chi^2$ cannot really distinguish the 3rd and 4th from the 1st and 2nd, as this is only information on distribution, not connected with randomness.

When we split the sequences and obtain $m$ different values $\lambda_n^m$ for each sequence, we can compute their mean value, $\lambda_n^m (mean)$.
Also we can define empiric distribution function $\Phi_n(\lambda)$ for $\lambda_n^m$ in similar way to \ref{eq:empiricdistribution}, and calculate the distance of that function from the theoretical one $\Phi(\lambda)$ - $\chi^2(\Phi_n(\lambda), \Phi(\lambda))$.

The following are the results of calculations with split of each sequence. There are two parameters, $|\lambda_n^m(mean)-\lambda_{mean}|$, where 
$$
\lambda_{mean}=\int{\lambda\phi(\lambda)d\lambda}\approx 0.875029,
$$ 
and 
$$
\chi^2(\Phi_n(\lambda), \Phi(\lambda)).
$$ 
Both parameters show that a sequence is closer to a random one, the closer is their value to 0.

\begin{figure}[!htbp]
	\centering
		\includegraphics[width=2.3 in]{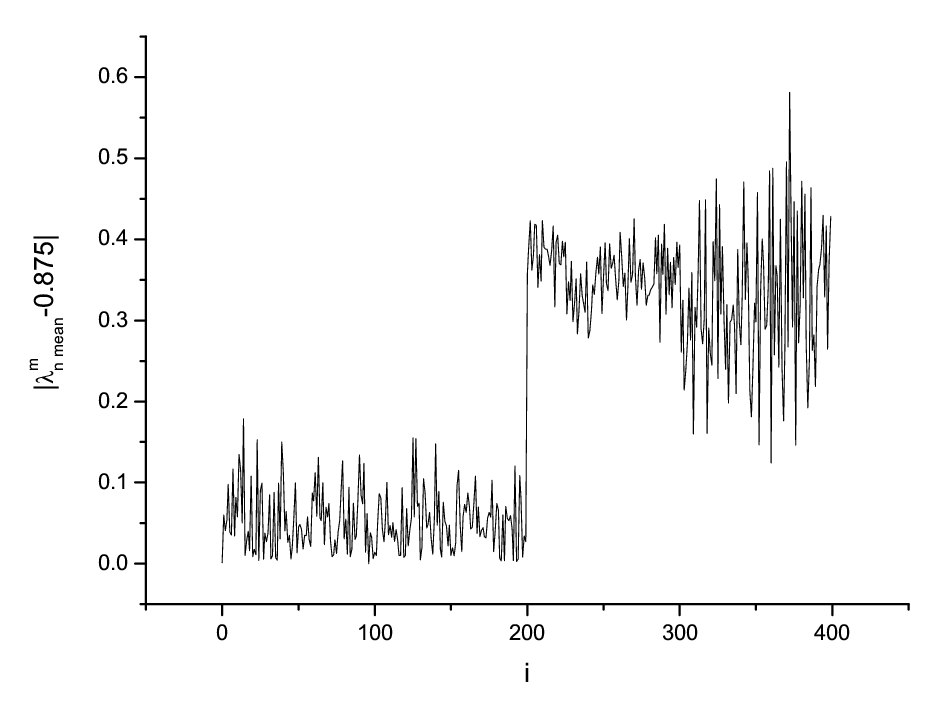}
		\includegraphics[width=2.3 in]{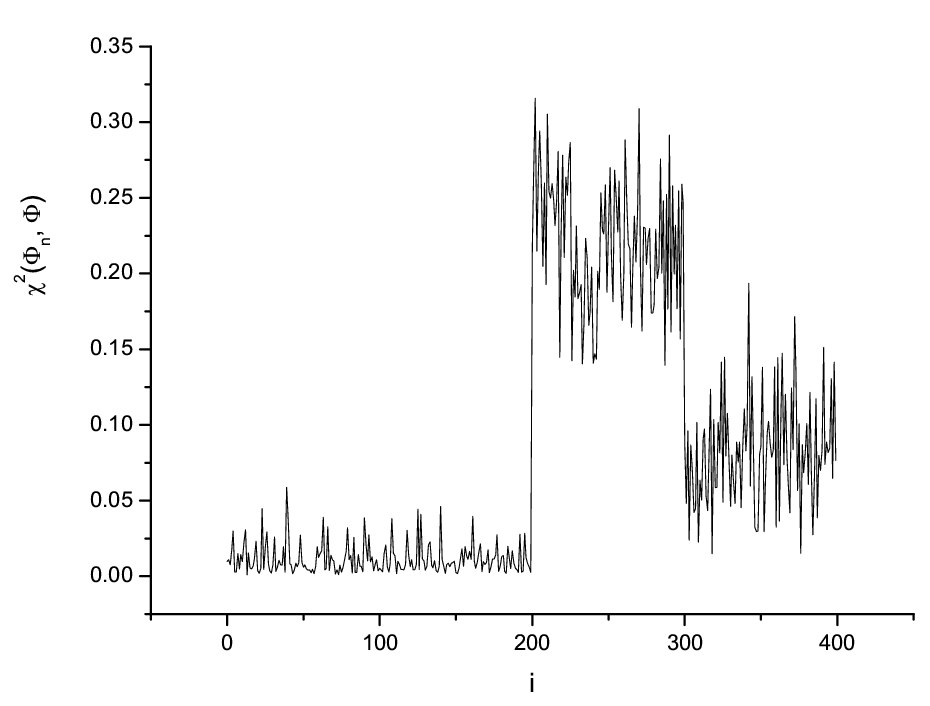}
	\caption{The behavior of $|\lambda_n^m(mean)-0.875029|$ and $\chi^2(\Phi_n(\lambda), \Phi(\lambda))$.}
	\label{fig:test_2}
\end{figure}

Figure \ref{fig:test_2} shows that the sequence is stochastic if 
$$
|\lambda_n^m(mean)-0.875029|<0.18
$$ 
or if 
$$
\chi^2(\Phi\textsl{}_n(\lambda), \Phi(\lambda))<0.03,
$$ 
and is non-stochastic otherwise.

These two criteria do not distinguish the 1st and 2nd groups from each other since we are dealing with small subsequences that do not "feel" the perturbation of distribution, however if $n$ is much bigger, then we could see their difference too.
Note also that these ranges can change depending on $n$, with larger values of $n$ the test will be more efficient, also there is a chance that they depend on theoretical distribution $F(x)$ of sequence $x_n$. This can be solved with consideration of $F(x_n)$ for the sequence with a uniform distribution. 

\section{Conclusions}

We studied datasets of various statistical properties to reveal the role of perturbations in the Cosmic Microwave Background, first of all due to the Galactic disk. We model different sequences of perturbed Gaussian distribution to obtain the behavior of the Kolmogorov's stochasticity parameter. This revealed the differences in the K-maps for CMB W and V bands of WMAP 5-year data, which are differently influenced by the contribution of the Galactic disk. Particularly, the Galactic radiation possess high value of $\lambda$ and $\Phi$ around 1, which shows its non-Gaussian nature, while CMB signal is basically Gaussian with inhomogeneous perturbations of various degree of randomness. 

The performed analysis shows that the Kolmogorov's parameter can serve as an indicator for separation of the CMB signal from the noise due to their different statistics.  

We thank Profs. V.G.Gurzadyan and A.A.Kocharyan for discussions and help.


\begin{thebibliography}{99}

\bibitem{01}
\label{Kolm}
A.N.~Kolmogorov, G.Ist.Ital.Attuar, {\bf 4} (1933) 83.
  
\bibitem{02}
\label{Arnold}
V.~Arnold, Nonlinearity, {\bf 21} (2008) T109. 

\bibitem{03}
\label{Arnold1}
V.I.~Arnold, ICTP/2008/001, Trieste.

\bibitem{04}
\label{GK_KSP}
V.G.~Gurzadyan, A.A.~Kocharyan, A \& A {\bf 492}, L33 (2008)
  
\bibitem{05}\label{K_sky}
V.G.~Gurzadyan, A.E.Allahverdyan et al, 2009, A\&A (in press); arXiv:0811.2832

\bibitem{07}
\label{Cruz}
M. Cruz, E. Martinez-Gonzalez, P. Vielva, arXiv:0901.1986 

\bibitem{06}
\label{GK_void}
V.G.~Gurzadyan, A.A.~Kocharyan, A \& A, {\bf 493}, L61 (2009)

\bibitem{08}\label{Das_Sp}
S. Das, D.N. Spergel, arXiv:0809.4704.  

\bibitem{09}\label{Arnold_book}
V.I. Arnold, Mathematical Methods of Classical Mechanics, Springer-Verlag, 1989. 

\end{thebibliography}
\end{document}